\documentclass[aps,showpacs,superscriptadress,preprint]{revtex4}
\usepackage{graphicx}
\usepackage{amsmath}
\usepackage{amssymb}
\usepackage{mathrsfs}

\begin{document}
\title{Quasinormal modes of scalar field coupled to Einstein's tensor in the non-commutative geometry inspired black hole}
\author{Zening Yan$^{1}$, Chen Wu$^{2}$\footnote{Electronic address: wuchenoffd@gmail.com} and Wenjun Guo$^{1}$} \affiliation{
\small 1. University of Shanghai for Science and Technology, Shanghai 200093, China\\
\small 2. Shanghai Advanced Research Institute, Chinese Academy of Sciences, Shanghai 201210, China}

\begin{abstract}
We investigate the quasinormal modes (QNMs) of the  scalar field coupled to the Einstein's  tensor in the non-commutative geometry inspired black hole spacetime. 
It is found that the lapse function of the non-commutative black hole metric can be represented by a Kummer's confluent hypergeometric function, which can effectively solve the problem that the numerical results of the QNMs are sensitive to the model parameters and make the QNMs values more reliable.
We  make a careful analysis of  the scalar QNM frequencies  by using several numerical methods,  and find that the numerical results obtained by the new WKB method (the Pad\'e approximants) and the Mashhoon method (P$\ddot{\text{o}}$schl-Teller potential method) are quite different from those obtained by the asymptotic iterative method (AIM) and time-domain integration method when the non-commutative parameter $\theta$ and coupling parameter $\eta$ are large.
The most obvious difference is that the numerical results obtained by the AIM and the time-domain integration method appear a critical value $\eta_c$ with an increase of $\eta$, which leads to the dynamical instability. After carefully analyzing the numeral results, we conclude that the numerical results obtained by the AIM and the time-domain integration method are closer to the theoretical values than those obtained by the WKB method and the Mashhoon method, when the $\theta$ and $\eta$ are large.
Moreover, through a numerical fitting,  we obtain that the functional relationship between the threshold $\eta_c$ and the non-commutative parameter $\theta$ satisfies $\eta_{c}=a\theta^{b}+c$ for a fixed $l$ approximately.
We find that the stability of dynamics can be ensured in the $\eta<\eta_c(\theta, l)$ region.

\end{abstract}

 \maketitle

\section{Introduction}
The investigations concerning the interaction of black holes with various fields
around  give us the possibility of obtaining some information about the physics of black holes. One of these information could be obtained from  quasinormal modes (QNMs) which are characteristic of the background black hole spacetimes.
QNMs is defined as the complex solution of perturbed wave equation under certain boundary conditions, it dominates the later time (ringdown) of the response of a black hole to its external disturbance. Recent astrophysical interests in QNMs originated from their relevance to gravitational wave analysis.
The Advanced LIGO detectors observed a transient gravitational-wave signal determined to be the coalescence of two black holes, thereby launching the era of gravitational wave astronomy \cite{1602.03837, 1606.04856, 1710.05832, 1706.01812, 1709.09660}.
Furthermore, it is believed that QNMs is closely related to the AdS/CFT correspondence  in string theory and loop quantum gravity  \cite{hep-th/9711200, hep-th/9802150, hep-th/9905104, hep-th/0002230, hep-th/0101133, hep-th/0205052, hep-th/9909056, gr-qc/0105103, gr-qc/0106032, gr-qc/0301052, hep-th/0302026}.
The AdS/CFT correspondence says that quasinormal modes of the $(D+1)$-dimensional asymptotically AdS black hole are poles of the retarded Green's function in the dual conformal field theory in $D$ dimensions at strong coupling \cite{Konoplya:2011qq}.  The lowest quasinormal frequencies of black holes have a direct interpretation as dispersion relations of hydrodynamic excitations in the ultrarelativistic heavy ion collisions. In addition, it is suggested that the basic features of AdS/CFT correspondence remain intact under non-commutative geometry \cite{Pramanik:2015eka}. All these motivated the  extensive numerical and analytical studies of QNMs for different spacetime and fields around black holes.

On the other hand, the scalar-tensor theory is one of the most popular alternative theory to modify the general relativity.
The most generic scalar-tensor model is described by the second-order field equation in four dimensions given by Horndeski Lagrangian \cite{Horndeski:1974wa}, one of the terms is the dynamic coupling of the scalar field and spacetime curvature. After that, the non-minimally coupled (NMC) model was first mentioned as an extension of the scalar-tensor theory \cite{Amendola:1990nn}. In this model, gravity may be regarded as a spontaneous symmetry-breaking effect \cite{Zee:1978wi, Accetta:1985du}. And the NMC term allows the existence of an oscillating universe \cite{RandjbarDaemi:1983jz}. Furthermore, the scalar-tensor theory is  extended to the non-minimal couplings between derivatives of a scalar field and the curvature terms, also known as the non-minimally derivative coupling (NMDC) model \cite{Amendola:1993uh}.

The general form of action with non-minimally derivative coupling between the scalar field and spacetime curvature can be represented as \cite{Gao:2010vr, Chen:2010qf}
\begin{equation}
\begin{aligned}
S= &\int d^{4} x \sqrt{-g} \bigg [\mathscr{F}\left(\Psi, \mathcal{R}, \mathcal{R}_{\mu \nu} \mathcal{R}^{\mu \nu}, \mathcal{R}_{\mu \nu \rho \sigma} \mathcal{R}^{\mu \nu \rho \sigma}\right)+\mathscr{K}\left(\Psi, \partial_{\mu} \Psi \partial^{\mu} \Psi, \nabla^{2} \Psi, \mathcal{R}^{\mu \nu} \partial_{\mu} \Psi \partial_{\nu} \Psi, \cdots\right)\\
   &+V(\Psi)\bigg ]+S_{(m)},
\end{aligned}
\end{equation}
where the nonlinear functions $\mathscr{F}$ merges all Lagrangian curvature terms and their coupling into scalar field components,
$\mathscr{K}$ denotes the general coupling between the curvature and the kinetic term of the scalar field, $V(\Psi)$ is the scalar field potential, and $S_{(m)}$ is the action of other matter fields.

Amendola \cite{Amendola:1993uh} has considered the most general gravity Lagrangian linear in the curvature scalar $\mathcal{R}$, quadratic in $\Psi$ and containing terms with four derivatives including all of the following terms:
\begin{equation}\begin{array}{lll}
\mathcal{L}_{1}=\kappa_{1} \mathcal{R} \Psi_{, \mu} \Psi^{, \mu}; \;\;\;\; & \mathcal{L}_{2}=\kappa_{2} \mathcal{R}_{\mu \nu} \Psi^{, \mu} \Psi^{, \nu}; \;\;\;\; & \mathcal{L}_{3}=\kappa_{3} \mathcal{R} \Psi \square \Psi; \\
\mathcal{L}_{4}=\kappa_{4} \mathcal{R}_{\mu \nu} \Psi \Psi^{; \mu \nu}; \;\;\;\; & \mathcal{L}_{5}=\kappa_{5} \mathcal{R}_{; \mu} \Psi \Psi^{, \mu}; \;\;\;\; & \mathcal{L}_{6}=\kappa_{6} \square \mathcal{R} \Psi^{2}.
\end{array}\end{equation}
In addition, the author also analyzed the derivative coupling model with only one mathematical term $\mathcal{L}_{1}$, and obtained some analytical inflationary solutions \cite{Amendola:1993uh}. Capozziello et al. investigated the general model with two coupling terms $\mathcal{L}_{1}$ and $\mathcal{L}_{2}$ and found that the de Sitter spacetime is an attractor solution of the model if $4 \kappa_{1}+\kappa_{2}>0$ is satisfied \cite{Capozziello:1999xt, Capozziello:1999uwa}.
Daniel and Caldwell discussed the case with a derivative coupling term $\mathcal{L}_{2}$, and they concluded that under the condition of weak coupling, $\kappa_{2}$ is severely constrained by precision tests of general relativity \cite{Daniel:2007kk}.
Because of nonlinear coupling, the equation of motion of scalar field is no longer the second-order differential equation but the fourth-order differential equation. However, Sushkov \cite{Sushkov:2009hk} found that when the scalar field is coupled with the Einstein's tensor, the equation of motion can be simplified to a second-order differential equation, which  means that this theory is a ``good'' dynamical theory from the point of view of physics.

The NMDC models gives the exact solution of the hairy black hole in the scalar-tensor gravity theories \cite{Benkel:2016rlz, Babichev:2015rva, Cisterna:2014nua, Anabalon:2013oea, Minamitsuji:2013ura}.
However, in Galileon theories, in order for the scalar field to have a non-trivial profile and to be finite on the horizon, these solutions must comply with strict constraints.
This model is usually used to describe inflation \cite{Germani:2010gm, Sadjadi:2013na, Sadjadi:2013psa, Goodarzi:2016iht, Li:2019ncw, Sadjadi:2012zp, Yang:2015pga} and late acceleration \cite{Sushkov:2009hk, Granda:2009fh, Sadjadi:2010bz, Sadjadi:2013uza, Li:2017dwr, Harko:2016xip, Jawad:2019bka, Quiros:2019gbt, Saichaemchan:2017psl, Matsumoto:2017gnx, Gumjudpai:2016frh, Granda:2010ex}.
NMDC acts as a friction term in early inflationary cosmology \cite{Amendola:1993uh, Sushkov:2009hk, Saridakis:2010mf, Granda:2012tx, Germani:2010gm}, which improves the early inflationary model and helps to solve the dark matter problem.

The solution of the first black hole of NMDC can not avoid the singular behavior, and the scalar field blows up on the event horizon.
There are two ways to avoid this problem, one is to introduce the mass term of the scalar field to break its shift symmetry \cite{Kolyvaris:2011fk, Kolyvaris:2013zfa}.
Another method is to allow the scalar field to be time-dependent and keep its shift symmetry.
It is explained in Ref. \cite{Babichev:2013cya} that for static spherically symmetric spacetime, the scalar field is non-trivial and regular, if the scalar field time-dependent. Furthermore, it is shown that the asymptotically flat solution or de-Sitter solution is permissible, and the regular hairy black hole solution is given \cite{Charmousis:2014zaa, Anabalon:2013oea, Minamitsuji:2013ura, Cisterna:2014nua, Babichev:2015rva, Sotiriou:2014pfa, Benkel:2016rlz}.

The references cited in this paragraph  have studied the QNMs of black holes under NMDC scalar field perturbation.
Refs. \cite{Chen:2010qf, Konoplya:2018qov} studied the dynamical evolution of a scalar field coupled to Einstein's tensor in the Reissner-Nordstr$\ddot{\text{o}}$m black hole spacetime. 
In addition, \cite{Konoplya:2018qov} proved that the phenomenon of arbitrarily long-lived QNMs of a massive scalar field in the vicinity of a black hole is not an artifact of the test field approximation, but takes place also when the derivative coupling of a scalar field with the Einstein's tensor is taken into consideration. Ref. \cite{Fontana:2018fof} studied the dynamical behavior of a scalar field non-minimally coupled to Einstein's tensor and Ricci scalar in the pure de-Sitter spacetime, and a new type of gravitational instability has been found.
In Ref. \cite{Abdalla:2019irr}, the authors discovered the dynamical instability of a Reissner-Nordstr$\ddot{\text{o}}$m-AdS black hole under perturbations of a massive scalar field coupled to the Einstein's tensor. Ref. \cite{Dong:2017toi} calculated the spectrum for a massless test scalar coupled both minimally to the metric, and non-minimally to the gravitational scalar.
The scalar QNMs of various static and spherically symmetric black holes with derivative coupling
black hole quasinormal modes in the scalar-tensor theory with  scalar field derivative coupling to the Einstein's tensor were also studied in \cite{Yu:2018zqd, Minamitsuji:2014hha, Gao:2010vr, Rinaldi:2012vy, Zhang:2018fxj, Lin:2011zzd, Zhou:2013zwa}.

At present, it is believed that the non-commutativity of spacetime will affect the gravity theory, and lots of researchers have directly constructed the  non-commutative modified gravity theory in different ways, such as \cite{Groenewold:1946kp, Szabo:2006wx, Chamseddine:2000si, Marculescu:2008gw, Chamseddine:2002fd, Cardella:2002pb, Chamseddine:2003we, Calmet:2006iz}. Another claim on this problem is that the non-commutativity of spacetime only affects all interactions except gravity, which means that the geometric part  of the Einstein's field equation is kept unchanged, and the non-commutativity only modifies the energy-momentum part. In this way, non-commutativity affects gravity in an indirect way, such as the coordinate coherent state formalism \cite{Smailagic:2003rp}.
This view can be regarded as a hypothesis, and the non-commutative effect inspired by string theory supports this hypothesis to some extent \cite{Nicolini:2005vd}.
However, N. Seiberg and E. Witten \cite{Seiberg:1999vs} proposed that the string theory background of non-commutative spacetime holds that gravity does not need to be modified.  On this basis, this paper discusses the scalar-tensor  theory in the background of the non-commutative spacetime.
Assuming that the non-commutativity of spacetime does not modify the Einstein-Hilbert action $S_{\textsl{E-H}}$ directly, the   spacetime non-commutativity is encoded in $g_{\mu \nu}$ that is a non-commutative inspired black hole solution obtained by the coordinate coherent state method.

The main purpose of this work is  listed as follows:
1) It is well known that  since the non-commutative black hole spacetime is not solution of the vacuum Einstein's equation, spherically symmetric Einstein's tensor $G^{\mu\nu}$ does not vanishes.
Then the effect of non-minimally derivative coupling on the QNMs of non-commutative black hole is not negligible \cite{Yu:2018zqd};
2) The coupling constant threshold $\eta_{c}$ as a function of the multipole number $l$   is obtained by fitting in Ref. \cite{Chen:2010qf}.
It is worth mentioning that  the independent variable $l$ is a discrete variable rather than continuous variable.
Therefore, it is more natural to  fit the $\eta_{c}$  as  a function of a continuous variable $\theta$ while making the $l$ as the  parameter of the fitting function;
3) Ref. \cite{Das:2018fzc} shows that there exist deviations and errors when one calculate the scalar QNMs in the non-commutative black hole spacetime by using the WKB method, and the higher-order results are not convergent. Therefore, in order to clarify this problem more transparently, we give our numerical study in the fourth part.

The rest of this paper is organized as follows.
In Section 2, we introduce the wavelike equation with positive and negative coupling parameter  in spherically symmetric spacetime.
In Section 3, the effective potential of the scalar field coupled with the Einstein's tensor in the non-commutative geometry inspired black hole spacetime is analyzed.  In Section 4, we use the WKB, the AIM and the Mashhoon method to calculate QNMs and compare these numerical results.
In Section 5, we investigate the dynamical evolution of scalar field coupled with the Einstein's tensor in the non-commutative geometry inspired black hole spacetime by time-domain integration method, and study the approximate equation of the $\eta$ threshold.
Finally, a brief summary of the full text is presented.

\section{Equation of motion of non-minimally derivative coupled scalar field in spherically symmetric spacetime}
We consider the action of the part of the massive scalar field coupled to the Einstein's tensor in the non-minimally derivative coupling model \cite{Sushkov:2009hk, MohseniSadjadi:2019nkw, Zhu:2015lry}, which can be written as
\begin{equation}
\begin{aligned}
S \big[g^{\mu \nu}, \Psi \big]  &= \int d^{4} x \sqrt{-g}\bigg \{ - \frac{1}{2}\big [\Psi_{,\mu} \Psi^{,\mu}+\kappa_{1} \mathcal{R} \Psi_{,\mu} \Psi^{,\mu}+\kappa_{2} \mathcal{R}^{\mu \nu}\Psi_{,\mu} \Psi_{,\nu}\big ]-\frac{1}{2}m^{2}\Psi^{2}\bigg \} \\
   &= \int d^{4} x \sqrt{-g}\bigg \{ - \frac{1}{2}\big [g^{\mu \nu}+\left(\kappa_{1} g^{\mu \nu} \mathcal{R}+\kappa_{2} \mathcal{R}^{\mu \nu}\right)\big ]\Psi_{,\mu} \Psi_{,\nu}-\frac{1}{2}m^{2}\Psi^{2}\bigg \},\\
\end{aligned}
\end{equation}
where the coupling parameters $\kappa_{1}$ and $\kappa_{1}$ are chosen as $-2 \kappa_{1}=\kappa_{2}=\kappa=\pm\eta$ \cite{Sushkov:2009hk}, and define $h^{\mu \nu} = g^{\mu \nu} \pm \eta G^{\mu \nu}$ \cite{Abdalla:2019irr, Fontana:2018fof} in this paper, obviously $G^{\mu \nu}= \mathcal{R}^{\mu v}-\frac{1}{2} g^{\mu v} \mathcal{R} $ is the Einstein's tensor.
\begin{equation}\label{action}
\begin{aligned}
S \big[g^{\mu \nu}, \Psi \big]  &= \int d^{4} x \sqrt{-g}\bigg \{ - \frac{1}{2}\big [g^{\mu \nu} + \kappa G^{\mu \nu}\big ]\Psi_{,\mu} \Psi_{,\nu}-\frac{1}{2}m^{2}\Psi^{2}\bigg \} \\
   &= \int d^{4} x \sqrt{-g}\bigg \{ - \frac{1}{2}\big [g^{\mu \nu} \pm \eta G^{\mu \nu}\big ]\Psi_{,\mu} \Psi_{,\nu}-\frac{1}{2}m^{2}\Psi^{2}\bigg \} \\
   &= \int d^{4} x \sqrt{-g}\bigg \{ - \frac{1}{2} h^{\mu \nu} \Psi_{,\mu} \Psi_{,\nu}-\frac{1}{2}m^{2}\Psi^{2}\bigg \}. \\
\end{aligned}
\end{equation}
The equation of motion of the scalar field derived from the action (\ref{action}) is given by
\begin{equation}
\frac{1}{\sqrt{-g}} \partial_{\mu}\bigg [ \sqrt{-g}\Big (g^{\mu \nu}\pm\eta G^{\mu \nu}\Big )\partial_{\nu} \Psi \bigg ] - m^{2}\Psi=0,
\end{equation}
 $\eta$ is the non-minimally derivative coupling parameter, $+\eta$ is taken in \cite{Fontana:2018fof, Gao:2010vr, Zhou:2013zwa} and $-\eta$ in \cite{Chen:2010qf, Abdalla:2019irr, Konoplya:2018qov, Yu:2018zqd, Zhang:2018fxj, Dong:2017toi, Rinaldi:2012vy, Minamitsuji:2014hha}. In \cite{Sushkov:2009hk}, both cases are discussed at the cosmological level. This paper stipulates that the value of the symbol ``$\eta$'' is positive.

The spherical metric for the four-dimensional static spacetime is given by
\begin{equation}
d s^{2}=-f(r) d t^{2}+\frac{1}{f(r)} d r^{2}+  r^{2} d \Omega^{2},
\end{equation}
where $d \Omega^{2}=d \theta^{2}+\sin ^{2} \theta d \phi^{2}$.
Ansatz to separate variables form
\begin{equation}
\Psi(t, r, \theta, \phi)=\sum_{l, m }\mathrm{e}^{-i \omega t} b(r) R(r)  Y_{l, m }(\theta, \phi),
\end{equation}
and put it into  Eq. (\ref{action}), we can  obtain
\begin{equation}\label{waveequ1}
-\omega^{2} h^{00}(R b)+h^{11}(R b)^{\prime \prime}+\frac{1}{r^{2}}\left(r^{2} h^{11}\right)^{\prime}(R b)^{\prime}-\big [m^{2}+l(l+1) h^{22}\big ](R b)=0.
\end{equation}
 And the spherically symmetric Einstein's tensor $G^{\mu \nu}$ is given by
\begin{equation}
G^{\mu \nu}=\left(\begin{array}{cccc}
\frac{A(r)}{f} & & & \\
& - A(r)f & & \\
& & \frac{B(r)}{r^{2}} & \\
& & & \frac{B(r)}{r^{2} \sin ^{2} \theta}
\end{array}\right),
\end{equation}
where
\begin{equation}
A(r)=\frac{1-f}{r^{2}}-\frac{f^{\prime}}{r}, \;\;\;\; B(r)=A(r)-\frac{\mathcal{R}}{2} =\frac{f^{\prime}}{r}+\frac{f^{\prime \prime}}{2},
\end{equation}
\begin{equation}
\mathcal{R}\equiv -\left[f^{\prime \prime}+\frac{4 f^{\prime}}{r}+\frac{2(f-1)}{r^{2}}\right].
\end{equation}
Dividing Eq. (\ref{waveequ1}) by $(-h^{00})$, and substituting 
\begin{equation}
h^{00}=g^{00} \pm \eta G^{00} = -\frac{1}{f} \pm \frac{\eta A}{f}, \; h^{11}=g^{11} \pm \eta G^{11} = f \mp \eta A f, \; h^{22}=g^{22} \pm \eta G^{22}=\frac{1}{r^{2}} \pm \frac{\eta B}{r^{2}},
\end{equation}
into Eq. (\ref{waveequ1}), we can obtain
\begin{equation}
\omega^{2}(R b)+f^{2} (R b)^{\prime \prime}+\Delta(r)_{\pm} (R b)^{\prime}-\Theta(r)_{\pm} (R b)=0,
\end{equation}
where
\begin{equation}
\Delta(r)_{\pm}=-\frac{\left(r^{2} h^{11}\right)^{\prime}}{r^{2} h^{00}}=f^{2}\left(\frac{\mp \eta A^{\prime}}{1 \mp \eta A}+\frac{f^{\prime}}{f}+\frac{2}{r}\right),
\end{equation}
\begin{equation}
\Theta(r)_{\pm}=-\frac{\big [l(l+1) h^{22}+m^{2}\big ]}{h^{00}}=\frac{f}{1 \mp \eta A}\left[(1 \pm \eta B) \frac{l(l+1)}{r^{2}}+m^{2}\right],
\end{equation}
and the subscript $\pm$ corresponds to the positive coupling parameter $+\eta$ and the negative coupling parameter $-\eta$, respectively.
Applying the tortoise coordinate $d r_{*}= d r/f$ and then defining
\begin{equation}
b(r)_{\pm}=\frac{1}{r \sqrt{1 \mp \eta A(r)}},
\end{equation}
we get  standard form of the Schr$\ddot{\text{o}}$dinger-like wave equation as
\begin{equation}\label{waveequ}
\frac{\partial^{2} {R}}{\partial r_{*}^{2}}+\Big[\omega^{2}- V(r)_{\pm}\Big] {R}=0,
\end{equation}
where the effective potential $V(r)_{\pm}$ reads
\begin{equation}
V(r)_{\pm}=\left[\Theta-\left(\frac{f^{2} b^{\prime \prime}+\Delta b^{\prime}}{b}\right)\right]_{\pm}.
\end{equation}
Therefore, the potential $V(r)_{+}$ can be written as
\begin{equation}\label{Vpositive}
\begin{aligned}
V(r)_{+} =& \frac{f}{1 - \eta A}\left[\frac{l(l+1)}{r^{2}}(1 + \eta B)+m^{2}\right]+\\
              &  f^{2}\left[\frac{-\eta}{1-\eta A}\left(\frac{A^{\prime \prime}}{2}+\frac{A^{\prime} f^{\prime}}{2 f}+\frac{A^{\prime}}{r}\right)+\frac{f^{\prime}}{r f}-\frac{1}{4}\left(\frac{-\eta A^{\prime}}{1-\eta A}\right)^{2}\right],\\
\end{aligned}
\end{equation}
and the potential $V(r)_{-}$  as
\begin{equation}\label{Vnegative}
\begin{aligned}
V(r)_{-} =& \frac{f}{1 + \eta A}\left[\frac{l(l+1)}{r^{2}}(1 - \eta B)+m^{2}\right]+\\
              &  f^{2}\left[\frac{+\eta}{1+\eta A}\left(\frac{A^{\prime \prime}}{2}+\frac{A^{\prime} f^{\prime}}{2 f}+\frac{A^{\prime}}{r}\right)+\frac{f^{\prime}}{r f}-\frac{1}{4}\left(\frac{+\eta A^{\prime}}{1+\eta A}\right)^{2}\right].\\
\end{aligned}
\end{equation}


\section{non-commutative geometry inspired black hole}
The lapse function of non-commutative geometry inspired black hole \cite{Nicolini:2005vd} is
\begin{equation}\label{metric1}
f(r)=1-\frac{4 M}{r \sqrt{\pi}} \gamma\left(\frac{3}{2}, \frac{r^{2}}{4 \theta}\right),
\end{equation}
which is the solution in the case of  four-dimensional Gaussian smeared matter distribution in non-commutative spacetime.
By means of  the equation
\begin{equation}
\gamma\left(\frac{3}{2}, \frac{r^{2}}{4 \theta}\right)=\frac{\sqrt{\pi}}{2}-\Gamma\left(\frac{3}{2}, \frac{r^{2}}{4 \theta}\right),
\end{equation}
the lapse function (\ref{metric1}) can be written as  the sum of the Schwarzschild part and upper incomplete gamma function part as follows
\begin{equation}\label{metric2}
f(r)=1-\frac{2 M}{r}+\frac{4 M}{r \sqrt{\pi}} \Gamma\left(\frac{3}{2}, \frac{r^{2}}{4 \theta}\right).
\end{equation}
When the condition $r^{2} \gg4 \theta$ is satisfied, Eq. (\ref{metric2}) is reduced to the Schwarzschild metric.
According to the recursive relation and the definition of Error function,
\begin{equation}
\gamma(s+1, z)=s \gamma(s, z)-z^{s} \mathrm{e}^{-z}, \;\;\;\; \gamma\left(\frac{1}{2}, z\right)=\sqrt{\pi} \mathrm{Erf}(\sqrt{z}),
\end{equation}
the lapse function (\ref{metric1}) can also described as
\begin{equation}\label{metric3}
f(r)=1-\frac{2 M}{r} \mathrm{Erf}\left(\frac{r}{2 \sqrt{\theta}}\right)+\frac{2 M}{\sqrt{\pi \theta}} \mathrm{e}^{-\frac{r^{2}}{4 \theta}}.
\end{equation}
By virtue of
\begin{equation}
\gamma(s, z)=\sum_{k=0}^{\infty} \frac{(-1)^{k}}{k !} \frac{z^{s+k}}{s+k}=s^{-1} z^{s} \mathrm{e}^{-z} \mathcal{M}\big (1, s+1, z \big ) =\frac{z^{s}}{s} \mathcal{M}\big (s, s+1,-z \big),
\end{equation}
where $\mathcal{M}$ is the Kummer confluent hypergeometric function.
Therefore, lapse function (\ref{metric1}) can be written as
\begin{equation}\label{metric4}
f(r)=1 - \frac{M r^2}{3 \sqrt{\pi} {\theta}^{3/2}} \mathcal{M}\bigg(\frac{3}{2}, \frac{5}{2},-\frac{r^{2}}{4 \theta} \bigg).
\end{equation}
We summarize these equivalent forms in Table. \ref{lapse}.
It can be seen that the lapse function of the non-commutative geometry inspired black hole has no singularity, that is, $\lim\limits_{r \rightarrow 0} f(r)_{NC}=1$, where the Eq. (\ref{metric4}) is more intuitive.

When using lapse function (\ref{metric1}) to calculate the QNMs of a non-commutative black hole, there will be a problem that the numerical results are sensitive to  the model parameters (peak potential $V_{\max }$ corresponds to $\widetilde{r}$ and the outer event horizon ${r}_2$), that is, the numerical results will fluctuate dramatically when different the model parameters are given \cite{Yan:2020hga}.
Choosing the expression form of Kummer confluent hypergeometric function can solve the problem that QNMs is affected by model parameters in practical calculation, and more accurate numerical results can be obtained, especially when the $\theta$ is small.
In addition, the numerical results calculated by AIM and WKB method with Kummer confluent hypergeometric function are in good agreement, which shows that choosing lapse function (\ref{metric4}) has a great advantage in practical calculation. As shown in Fig. 2.

\begin{table}[tbh]\centering
\caption{Several equivalent forms of the lapse function of non-commutative geometry inspired black hole.
\vspace{0.3cm}} \label{lapse}
\begin{tabular*}{16cm}{*{2}{l @{\extracolsep\fill}}}
\hline
\hline
Lapse function & Main function form \\
\hline
$f(r)=1-\frac{4 M}{r \sqrt{\pi}} \gamma(\frac{3}{2}, \frac{r^{2}}{4 \theta})$    &  Lower incomplete gamma function: $\gamma$\\
$f(r)=1-\frac{2 M}{r}+\frac{4 M}{r \sqrt{\pi}} \Gamma(\frac{3}{2}, \frac{r^{2}}{4 \theta})$	    & Upper incomplete gamma function: $\Gamma$\\
$f(r)=1-\frac{2 M}{r} \mathrm{Erf}(\frac{r}{2 \sqrt{\theta}})+\frac{2 M}{\sqrt{\pi \theta}} \mathrm{e}^{-\frac{r^{2}}{4 \theta}}$   & Gauss error function: $\mathrm{Erf}$ \\
$f(r)=1 - \frac{M r^2}{3 \sqrt{\pi} {\theta}^{3/2}} \mathcal{M}(\frac{3}{2}, \frac{5}{2},-\frac{r^{2}}{4 \theta})$        & Kummer confluent hypergeometric function: $\mathcal{M}$ \\
\hline
\hline
\end{tabular*}
\end{table}

In addition, the condition for the existence of the event horizon of this non-commutative geometry inspired black hole: the range of non-commutative parameters is $0<\theta\lesssim0.275811$ \cite{Nicolini:2005vd, Yan:2020hga}, when $M=1$ is selected.

Substituting (\ref{metric1}) into (\ref{Vpositive}) (\ref{Vnegative}) respectively, we obtain
\begin{equation}
V(r)_{+}=\frac{\mathscr{P}(r)}{16 \pi\theta^{\frac{5}{2}} r^5  \left(\sqrt{\pi} \theta \mathrm{e}^{\frac{r^2}{4 \theta}}-\frac{\eta M}{\sqrt{\theta}}\right)^2},
\end{equation}
\begin{equation}
V(r)_{-}=\frac{\mathscr{N}(r)}{16 \pi\theta^{\frac{5}{2}} r^5  \left(\sqrt{\pi} \theta \mathrm{e}^{\frac{r^2}{4 \theta}}+\frac{\eta M}{\sqrt{\theta}}\right)^2},
\end{equation}
where $\mathscr{P}(r)$, $\mathscr{N}(r)$ is a function of parameters $\{M,\theta,\eta,m,l\}$, respectively.
It can be seen that the equation $\sqrt{\pi} \theta \mathrm{e}^{\frac{r^2}{4 \theta}}-{\eta M}/{\sqrt{\theta}}=0$ must have a real root at $r>0$, which means the potential $V(r)_{+}$ has singularity. However, the potential $V(r)_{-}$ does not behave singularity at $r>0$. Therefore, this paper only discusses the negative coupling potential $V(r)_{-}$.
\begin{figure}[htbp]
\centering
\includegraphics[height=10.6cm,width=16cm]{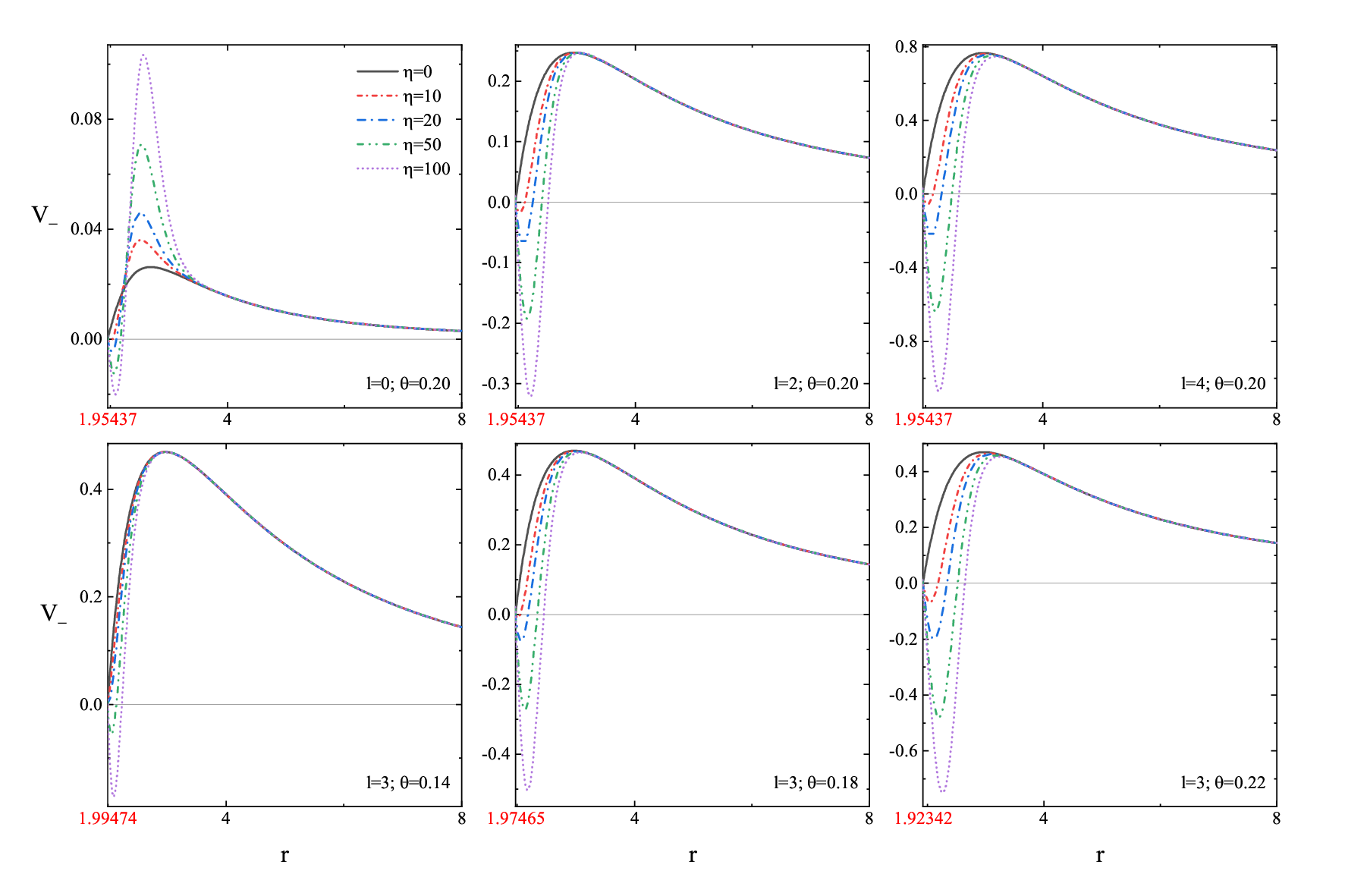}
\caption{Variation of the effective potential  $V(r)_{-}$ with respect to the radial coordinate $r$ for different values of $\theta$ or $\eta$.}
\end{figure}

We show the effective potential $V(r)_{-}$ as a function of $r$ for different values of  $\theta$ and $\eta$ in Fig. 1.
The following information can be seen in the figure:
1) When $l=0$, the peak value of effective potential $V(r)_{-}$ increases with the increase of $\eta$, which is contrary to other values of $l$.
2) When the $\theta$ value is small, the peak value of $V(r)_{-}$ hardly changes with the increase of $\eta$. 
3) When the values $l$ and $\theta$ increase, the negative region of V increases.

\section{Calculation methods and numerical results}
\subsection{Introduction of the WKB, the AIM and the Mashhoon method}
The real part of QNMs corresponds to the oscillation frequency, while the imaginary part corresponds to the damping.
Therefore, the complex $\omega$ values are written as $\omega = \text{Re}(\omega) + i \text{Im}(\omega)$.
There are different numerical methods for calculating QNMs.
In this paper, the QNMs of a scalar field coupling to Einstein's tensor in the background of non-commutative geometry inspired black hole is calculated by using the WKB method \cite{Schutz:1985zz}, the AIM \cite{Ciftci:2005xn} and the Mashhoon method \cite{Poschl:1933zz}.
Next we give a brief introduction to these  methods.

The WKB approximation method was applied for the first time by Schutz and Will \cite{Schutz:1985zz}. Iyer and his coworkers developed the WKB method up to 3rd \cite{Iyer:1986np} order and later, Konoplya developed it up to 6th order \cite{Konoplya:2003ii}.
Reference \cite{Matyjasek:2017psv} uses Pad\'e approximants to help guess the asymptotic behavior of WKB series, which greatly increases the accuracy of WKB method. And the latest WKB method has been extended to 13th order approximation.
This semi-analytical method has been applied extensively in numerous black hole spacetime cases, which has been proved to be accurate up to around one percent for the real and the imaginary parts of the quasinormal frequencies for low-lying modes with $n < l$, where $n$ is the mode number and $l$ is the angular momentum quantum number.

The N order formalism  of the WKB approximation has formula
\begin{equation}
\frac{iQ_{0}}{\sqrt{-2 Q_{0}^{\prime \prime}}}-\sum_{k=2}^{N} \Lambda_{k}=n+\frac{1}{2},
\end{equation}
where the correction term $\Lambda_{k}$ gives different orders in \cite{Konoplya:2003ii, Matyjasek:2017psv}.
$Q_{0}^{(i)}$ represents the $i$-th derivative of $Q=\omega^{2}-V$ at its maximum tortoise coordinate $r_{+}$.

In Ref. \cite{Ciftci:2005xn}, the asymptotic iterative method (AIM) was applied to solve second order differential equations for the first time.
This new method was then used to obtain the QNM frequencies of field perturbation in Schwarzschild black hole spacetime \cite{Cho:2009cj} and other black hole spacetime \cite{Ponglertsakul:2020ufm}.
Let's consider a second order differential equation of the form
\begin{equation}\label{2nd Eq}
\chi''=\lambda_{0}(x)\chi'+s_{0}(x)\chi,
\end{equation}
where $\lambda_{0}(x)$ and $s_{0}(x)$ are well defined functions and sufficiently smooth. Differentiating the equation above with respect to $x$  leads to
\begin{equation}
\chi'''=\lambda_{1}(x)\chi'+s_{1}(x)\chi,
\end{equation}
where the two coefficients are  $\lambda_{1} (x) =\lambda_{0}'+s_{0}+\lambda_{0}^{2}$ and $s_{1}(x)=s_{0}'+s_{0}\lambda_{0}.$
Using this process iteratively, differentiate $n$ times with respect to the independent variable, which produces the following equation
\begin{equation}
\chi^{(n+2)}=\lambda_{n}(x)\chi'+s_{n}(x)\chi,
\end{equation}
where the new coefficients $\lambda_{n}(x)$ and $s_{n}(x)$ are associated with the older ones through the following relation
\begin{equation}
\lambda_{n}(x)=\lambda'_{n-1}(x)+s_{n-1}(x)+\lambda_{0}(x)\lambda_{n-1}(x), \;\;\;\; s_{n}(x)=s'_{n-1}(x)+s_{0}(x)\lambda_{n-1}(x),\label{iteration}
\end{equation}
for sufficiently large values of $n$, the asymptotic concept of the AIM method is introduced by \cite{Cho:2011sf}
\begin{equation}
\frac{s_{n}(x)}{\lambda_{n}(x)}=\frac{s_{n-1}(x)}{\lambda_{n-1}(x)} = \text{Constant}. \label{Quantum condition}
\end{equation}

The perturbation frequency can be obtained from the above-mentioned "quantization condition". Then, $\lambda_{n}(x)$ and $s_{n}(x)$ are expanded into Taylor series around the point $x'$ at which the AIM method is performed
\begin{equation}
\lambda_{n}\left(x^{\prime}\right)=\sum_{i=0}^{\infty} c_{n}^{i}\left(x-x^{\prime}\right)^{i}, \;\;\;\; s_{n}\left(x^{\prime}\right)=\sum_{i=0}^{\infty} d_{n}^{i}\left(x-x^{\prime}\right)^{i},
\end{equation}
where $c_{n}^{i}$ and $d_{n}^{i}$ are the $i$-th Taylor coefficients of $\lambda_{n}(x')$ and $s_{n}(x')$, respectively. Substitution of above equations into Eq. (\ref{iteration}) leads to a set of recursion relations for the Taylor coefficients as
\begin{equation}
c_{n}^{i}=\sum_{k=0}^{i} c_{0}^{k} c_{n-1}^{i-k}+(i+1) c_{n-1}^{i+1}+d_{n-1}^{i}, \;\;\;\; d_{n}^{i}=\sum_{k=0}^{i} d_{0}^{k} c_{n-1}^{i-k}+(i+1) d_{n-1}^{i+1},
\label{recursion}
\end{equation}
after applying the recursive relation (\ref{recursion}) in Eq. (\ref{Quantum condition}), the quantization condition can be obtained
\begin{equation}
d_{n}^{0}c_{n-1}^{0}- c_{n}^{0} d_{n-1}^{0}=0,
\end{equation}
which can be used to calculate the QNMs of black holes more accurately.

Mashhoon method is the P$\ddot{\text{o}}$schl-Teller potential approximation method \cite{Poschl:1933zz}, also known as the inverted potential method (IPM), which uses the P$\ddot{\text{o}}$schl-Teller potential $V_{P T}$ to approximate the effective potential $V$ in the tortoise coordinate system
\begin{equation}
V_{P T}=\frac{V_{0}}{\cosh ^{2} \alpha\left(x -x_{0}\right)}, \quad -2 V_{0} \alpha^{2}=\left.\frac{d^{2} V}{d x^{2}}\right|_{x=x_{0}},
\end{equation}
where $V_{0}$ is the height of the effective potential and $-2 V_{0} \alpha^{2}$ is the curvature of the potential at its maximum.
The bound states of the P$\ddot{\text{o}}$schl-Teller potential are well known
\begin{equation}
\Omega=\alpha^{\prime}\left[-\left(n+\frac{1}{2}\right)+\left(\frac{1}{4}+\frac{V_{0}}{\left(\alpha^{\prime}\right)^{2}}\right)^{1 / 2}\right], \quad n=0,1,2,\cdots.
\end{equation}
The quasinormal modes $\omega$ can be obtained from the inverse transformation $\alpha^{\prime}=i \alpha$ as follows
\begin{equation}
\omega=\pm \sqrt{V_{0}-\frac{1}{4} \alpha^{2}}-i \alpha\left(n+\frac{1}{2}\right), \quad n=0,1,2,\cdots.
\end{equation}
It is well known that for the low-lying QNMs, in the majority of cases the behavior of the effective potential is essential only in some region near the black hole, so that the fit of the height of the effective potential and of its second derivative is indeed enough for calculation of QNM frequencies.
This method gives quite accurate estimation for the high multipole number modes.

As far as these numerical methods are concerned, we would like to give some comments related to our numerical discussion.
It is found that there are deviations and errors in  the calculation of  the QNMs of the non-commutative black hole by using the WKB method, so we intend to exploit other means to give more reasonable numerical results.
 Ref. \cite{Batic:2019zya} has compared the P$\ddot{\text{o}}$schl-Teller potential $V_{PT}$ with the effective potential $V_{eff}$ and
  concluded that the Mashhoon method can be used to calculate the QNMs of the non-commutative black hole, so we choose the Mashhoon method as one of our methods  in this context.  Meanwhile considering the inaccuracy of the Mashhoon method, we choose the AIM method for comparing  our numerical results.
  On the other hand, the Pad\'e approximants (a variant of the WKB method) is also applied in our calculations. The purpose of developing this method is to improve the accuracy of high-order WKB approximation, which is successfully applied in the case of the non-commutative black hole.  
  However, the mathematical foundation of this method has not been fully solved \cite{Matyjasek:2017psv, Konoplya:2019hlu}, so we should question the results obtained by this method and compare it with other methods.

\subsection{Comparison of numerical results}
Now,  we report the QNMs of non-minimally derivative coupled scalar field in the non-commutative geometry inspired black hole  by using the afore-mentioned methods.
For three $l$ values and $\eta=1$, we show the $Re(\omega)$ and $-Im(\omega)$ of QNMs as a function of $\theta$ obtained by the (6th, 5th, 4th, 3rd) order WKB approximation, the AIM method and the Mashhoon method in Fig. 2.  It can be seen that the  fluctuation of numerical results given by the 6th, 5th and 4th WKB method increase dramatically with the increase of $\theta$, and  the numerical results of the 3rd WKB method, the AIM method and the Mashhoon method are highly in good concordance.
In Refs. \cite{Das:2018fzc, Batic:2019zya}, it is mentioned that the higher-order WKB numerical results of non-commutative geometry inspired black hole  are not  convergent. Ref. \cite{Batic:2019zya} also points out that the Mashhoon method is more suitable for the non-commutative black hole spacetime.
\begin{figure}[htbp]
\centering
\includegraphics[height=8.2cm,width=16.4cm]{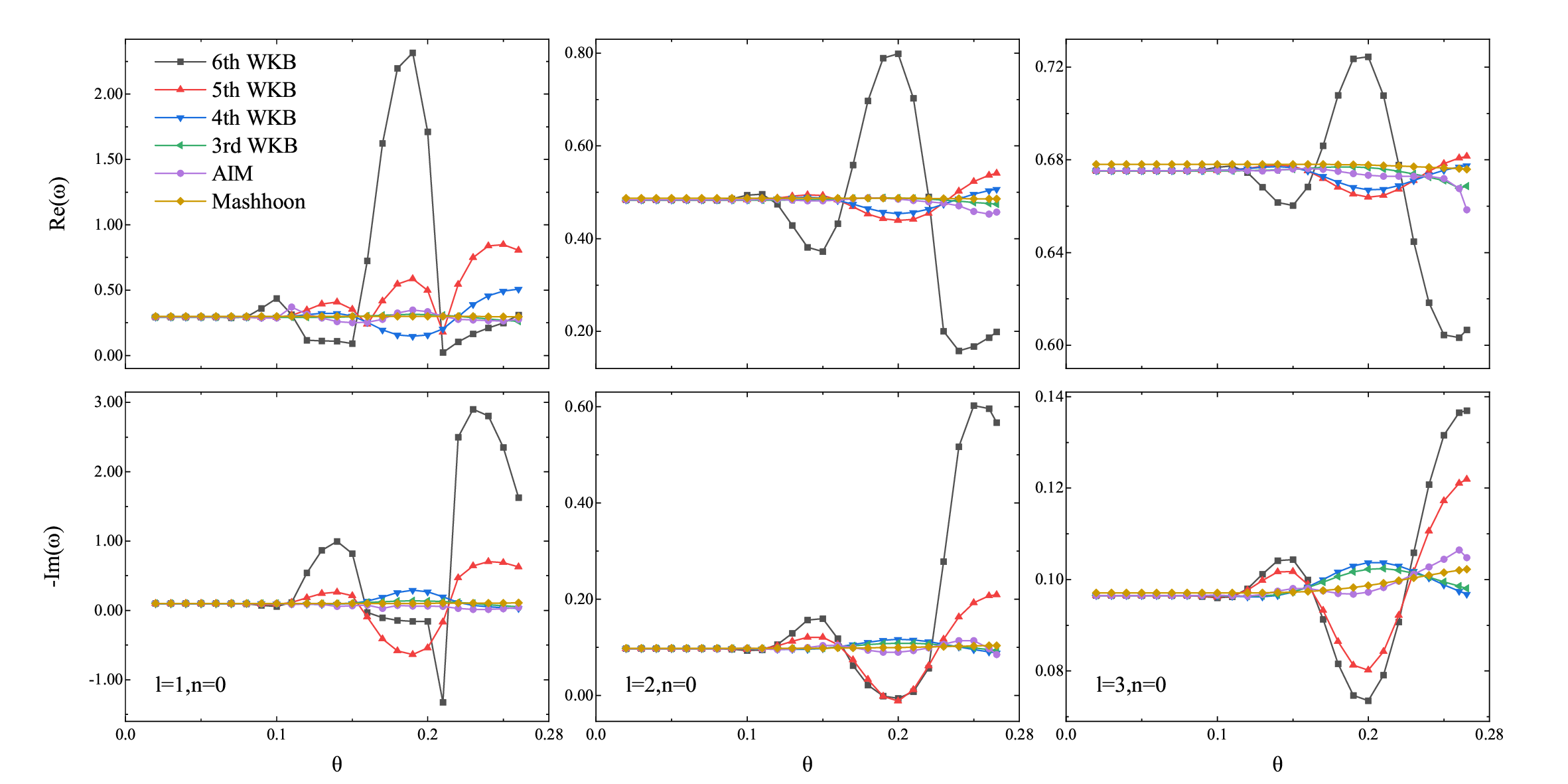}
\caption{$Re(\omega)$ and $-Im(\omega)$  parts of the QNMs with different $l$ values as a function of $\theta$ are  obtained by the (6th, 5th, 4th, 3rd) WKB method, the AIM method and  the Mashhoon method. The parameters are selected as $M=1$,$m=0$, $\eta=1$.}
\end{figure}

Next, we use the Pad\'e approximants to improve the accuracy of the WKB approximation and recalculate the QNMs.
For simplicity,  we compute the numerical results for the non-commutative parameter $\theta=0.02$ and $\theta=0.2$ to illustrate our discussion.

After using the Pad\'e approximants to improve the accuracy of the WKB method, when $\theta=0.02$, the absolute computational error of the 3rd $\sim$ 13th WKB numerical results is about $10^{-6} \sim 10^{-8}$. 
In this subsection, we show the accurate numerical results of two orders in the 3rd $\sim$ 13th order WKB approximation.
The numerical results of the 6th and 8th order WKB approximation are presented in Table. \ref{NMDC-NC-0.02}.
Therefore, when the $\theta$ value is small, the numerical results obtained by the Mashhoon method, the WKB method (the Pad\'e approximants) and the AIM method are  self-consistent.
\begin{table}[tbh]\centering
\caption{Comparison of the QNMs numerical results calculated by the Mashhoon method, the WKB method (the Pad\'e approximants) and the AIM method for different $\eta$ values when $\theta=0.02$. The parameters are selected as $M=1$, $l=3$, $n=0$, $m=0$ and the numerical results retain six significant digits.
\vspace{0.3cm}} \label{NMDC-NC-0.02}
\begin{tabular*}{16.4cm}{*{5}{c @{\extracolsep\fill}}}
\hline
$\eta$ & Mashhoon (P-T) & 6th WKB & 8th WKB & AIM \\
\hline
0   	 & 0.678098$-$0.0970906$i$ & 0.675366$-$0.0964997$i$ & 0.675366$-$0.0964997$i$ & 0.675366$-$0.0964996$i$ \\
1   	 & 0.678098$-$0.0970906$i$ & 0.675366$-$0.0964997$i$ & 0.675366$-$0.0964997$i$ & 0.675366$-$0.0964996$i$ \\
$\vdots$ & $\vdots$ & $\vdots$ & $\vdots$ & $\vdots$ \\
100 	 & 0.678098$-$0.0970906$i$ & 0.675366$-$0.0964997$i$ & 0.675366$-$0.0964996$i$ & 0.675366$-$0.0964996$i$ \\
\hline
\end{tabular*}
\end{table}

In the 3rd $\sim$ 13th order WKB approximation, when $\theta=0.2$, we show the numerical results of the 6th and 7th order.
For $\theta=0.2$, the numerical results of the 6th and 7th order WKB approximation are in good agreement with those of the AIM method and the Mashhoon method when the $\eta$ value is small, as shown in Table. \ref{NMDC-NC-0.2-S-eta}.
Therefore, when the $\theta$ value is quite large and the $\eta$ value is relatively small, the numerical results obtained by the Mashhoon method, the WKB method (the Pad\'e approximants) and the AIM method are in good agreement as well.
\begin{table}[tbh]\centering
\caption{Comparison of the QNMs numerical results calculated by the WKB method (the Pad\'e approximants), the Mashhoon method and the AIM method when $\eta$ value is small. The parameters are selected as $M=1$, $\theta=0.2$, $l=3$, $n=0$, $m=0$ and the numerical results is accurate to six decimal places.
\vspace{0.3cm}} \label{NMDC-NC-0.2-S-eta}
\begin{tabular*}{16.4cm}{*{4}{c @{\extracolsep\fill}}}
\hline
Method & $\omega$ $(\eta=1)$ & $\omega$ $(\eta=3)$ & $\omega$ $(\eta=5)$ \\
\hline
6th WKB	       & 0.674385 $-$ 0.099339$i$ & 0.674157 $-$ 0.104089$i$ & 0.675310 $-$ 0.108057$i$ \\
7th WKB        & 0.674401 $-$ 0.098854$i$ & 0.675363 $-$ 0.104256$i$ & 0.675479 $-$ 0.107941$i$ \\
Mashhoon (P-T) & 0.677697 $-$ 0.098692$i$ & 0.676846 $-$ 0.102063$i$ & 0.676074 $-$ 0.105003$i$ \\
AIM	           & 0.673311 $-$ 0.097236$i$ & 0.672692 $-$ 0.105613$i$ & 0.672112 $-$ 0.108950$i$ \\
\hline
\end{tabular*}
\end{table}

   The Table. \ref{NMDC-NC-0.2-L-eta} list the numerical results of the WKB method (the Pad\'e approximants), the Mashhoon method and  the AIM method for $\theta=0.2$ when the $\eta$ value is very large.  The imaginary parts of the WKB and Mashhoon results decrease with the increase of $\eta$, while the imaginary parts of the AIM results increase with the increase of $\eta$. From this table, we can conclude that the numerical results of the WKB method  are close to those of the Mashhoon method but different from those of the AIM method.
However, for large $\eta$ and $\theta$, the absolute computational error of the 3rd $\sim$ 13th order WKB numerical results is about $10^{-1} \sim 10^{-4}$, which means the accuracy of these numerical results is not reliable again.  
Considering that the accuracy of the Mashhoon method itself is not very high, we are not sure that the numerical results of the Mashhoon method and the WKB method (the Pad\'e approximants) are closer to theoretical values when the $\eta$ value is large, although these numerical results are close to each other.

In order to investigate the issue mentioned above more deeply, in the next section we will consider the dynamical evolution by using the time-domain integration method.

\begin{table}[tbh]\centering
\caption{Comparison of the QNMs numerical results calculated by the WKB method (the Pad\'e approximants), the  Mashhoon method and the AIM method when $\eta$ is large. The parameters are selected as $M=1$, $\theta=0.2$, $l=3$, $n=0$, $m=0$ and the numerical results is accurate to six decimal places.
\vspace{0.3cm}} \label{NMDC-NC-0.2-L-eta}
\begin{tabular*}{16.4cm}{*{4}{c @{\extracolsep\fill}}}
\hline
Method & $\omega$ $(\eta=10)$ & $\omega$ $(\eta=20)$ & $\omega$ $(\eta=40)$ \\
\hline
6th WKB	       & 0.674379 $-$ 0.114152$i$ & 0.673507 $-$ 0.123585$i$ & 0.671464 $-$ 0.133924$i$ \\
7th WKB        & 0.675087 $-$ 0.114673$i$ & 0.673674 $-$ 0.123443$i$ & 0.673936 $-$ 0.135204$i$ \\
Mashhoon (P-T) & 0.674400 $-$ 0.111049$i$ & 0.671759 $-$ 0.119846$i$ & 0.667955 $-$ 0.131308$i$ \\
AIM	           & 0.661346 $-$ 0.097077$i$ & 0.269445 $-$ 0.067265$i$ & 0.322828 $+$ 0.273602$i$ \\
\hline
\hline
Method & $\omega$ $(\eta=80)$ & $\omega$ $(\eta=90)$ & $\omega$ $(\eta=200)$ \\
\hline
6th WKB	       & 0.667905 $-$ 0.144677$i$ & 0.667067 $-$ 0.146754$i$ & 0.659920 $-$ 0.161092$i$ \\
7th WKB        & 0.669224 $-$ 0.145584$i$ & 0.668176 $-$ 0.147293$i$ & 0.659604 $-$ 0.158175$i$ \\
Mashhoon (P-T) & 0.662932 $-$ 0.144904$i$ & 0.661961 $-$ 0.147379$i$ & 0.654538 $-$ 0.165079$i$ \\
AIM	           & 0.414428 $+$ 0.643554$i$ & 0.430749 $+$ 0.707675$i$ & 0.541151 $+$ 1.157360$i$ \\
\hline
\end{tabular*}
\end{table}

\section{The evolution of perturbation is analyzed by the time-domain integration method}
\subsection{Time-domain integration method}
First of all, a brief introduction to the Gundlach-Price-Pullin method \cite{Gundlach:1993tp}, that is, the time-domain integration method or finite difference method.
In the time domain, we study the perturbation attenuation of scalar field coupling to Einstein's tensor corresponding to different $\eta$ in the background of a non-commutative geometry inspired black hole spacetime by using the numerical characteristic integral method,
that uses the light-cone variable $du=dt-dx$ and $dv=dt+dx$, and rewrite Eq. (\ref{waveequ}) as
\begin{equation}
-4 \frac{\partial^{2}}{\partial u \partial v} \Psi(u, v)=V_{i}\bigg[r\Big(\frac{v-u}{2}\Big)\bigg] \Psi(u, v).
\end{equation}
In the characteristic initial value problem, initial data are specified on the two null surfaces $u=u_{0}$ and $v=v_{0}$,
since the basic aspect of field attenuation has nothing to do with the initial conditions, it is assumed that the field $\Psi$ is initially in the form of Gaussian wave packets, so we choose the initial condition as $\Psi\left(u=u_{0}, v\right)=\exp \left[-\frac{\left(v-v_{c}\right)^{2}}{2 \sigma^{2}}\right]$, $\Psi\left(u, v=v_{0}\right)=0$, and choose the appropriate Gaussian wave package in the practical computation.
The discretization method we use is
\begin{equation}
\Psi(N)=\Psi(W)+\Psi(E)-\Psi(S)- \frac{\Delta^{2}}{8}\big [\Psi(W)+\Psi(E) \big]V(S)+\mathcal{O}\left(\Delta^{4}\right),
\end{equation}
where we have used the following definitions for the points: $N=(u+\Delta, v+\Delta)$, $W=(u+\Delta, v)$, $E=(u, v+\Delta)$ and $S=(u, v)$.
When the integration is completed, the value $\Psi\left(u_{\max }, v\right)$ is extracted, where $u_{\max }$ is the maximum value of $u$ on the numerical grid, as long as the $u_{\max }$ is large enough, we have a good approximation of the wave function at the event horizon.
In this way, we obtain the time-domain profile, which is a series of values of the perturbation field $\Psi(t=(v+u)/2, x=(v-u)/2)$ at a given position $x$ and discrete moments $t_{0}, t_{0}+h, t_{0}+2h, \cdots, t_{0}+Nh$.

In this paper, we  calculate the time-domain profiles of a scalar field coupling to Einstein's tensor in the background of the non-commutative geometry inspired black hole spacetime.
The Fig. 3 shows the dynamical evolution of the scalar field coupling to Einstein's tensor in the non-commutative geometry inspired black hole spacetime corresponding to different $l$ and $\eta$ values. We can find that the dynamical evolution always decays for arbitrary coupling parameter $\eta$ when the multipole number $l=0$.
There exists a critical coupling parameter $\eta_{c}$ when  $l>0$ so that the dynamical evolution  no longer decays  if $\eta > \eta_{c}$ is satisfied, which means that dynamical instability will occur. The Fig. 4 shows the dynamical evolution of the scalar field coupling to Einstein's tensor in the non-commutative geometry inspired black hole spacetime corresponding to different $l$ and $\theta$ values. Similarly, we can conclude that when the multipole number $l=0$, the dynamical evolution  always decays for all the $\theta$ values in the valid range.
When $l>0$, similarly, there will be a critical value $\theta_c$ so that dynamical instability will occur if $\theta > \theta_{c}$ takes place.

\begin{figure}[htbp]
\centering
\includegraphics[height=13.6cm,width=16.4cm]{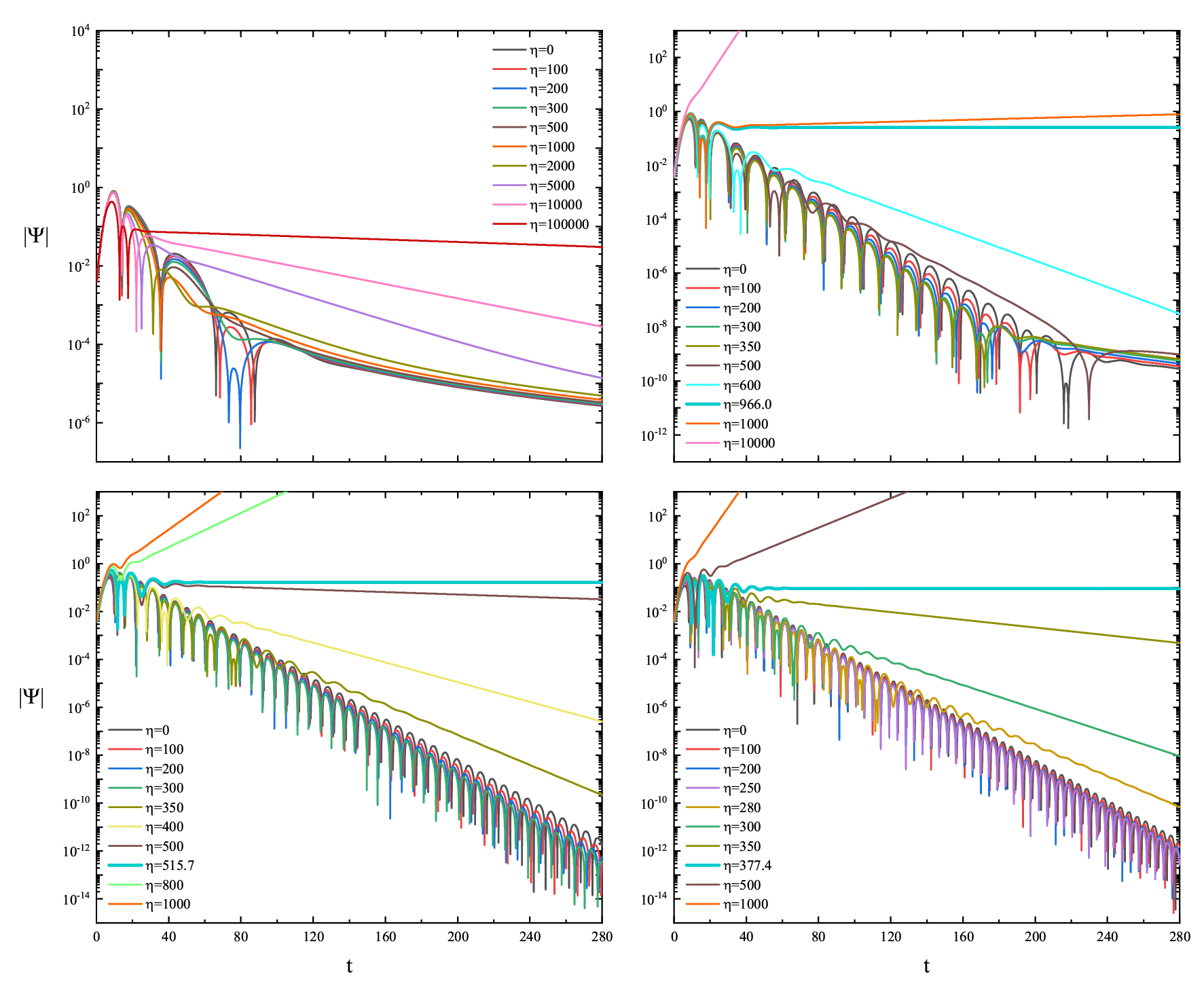}
\caption{Time-domain profiles of a scalar field coupling to Einstein's tensor corresponding to different $\eta$ in the background of non-commutative geometry inspired black hole spacetime. We set $\theta =0.1$ in all four panels. The figures from upper-left  to lower-right are corresponding to  $l$ = 0, 1, 2, and 3, respectively. The parameters are selected as $M=1$, $m=0$, $\sigma=3$, $v_{c}=10$.}
\end{figure}

\begin{figure}[htbp]
\centering
\includegraphics[height=13.6cm,width=16.4cm]{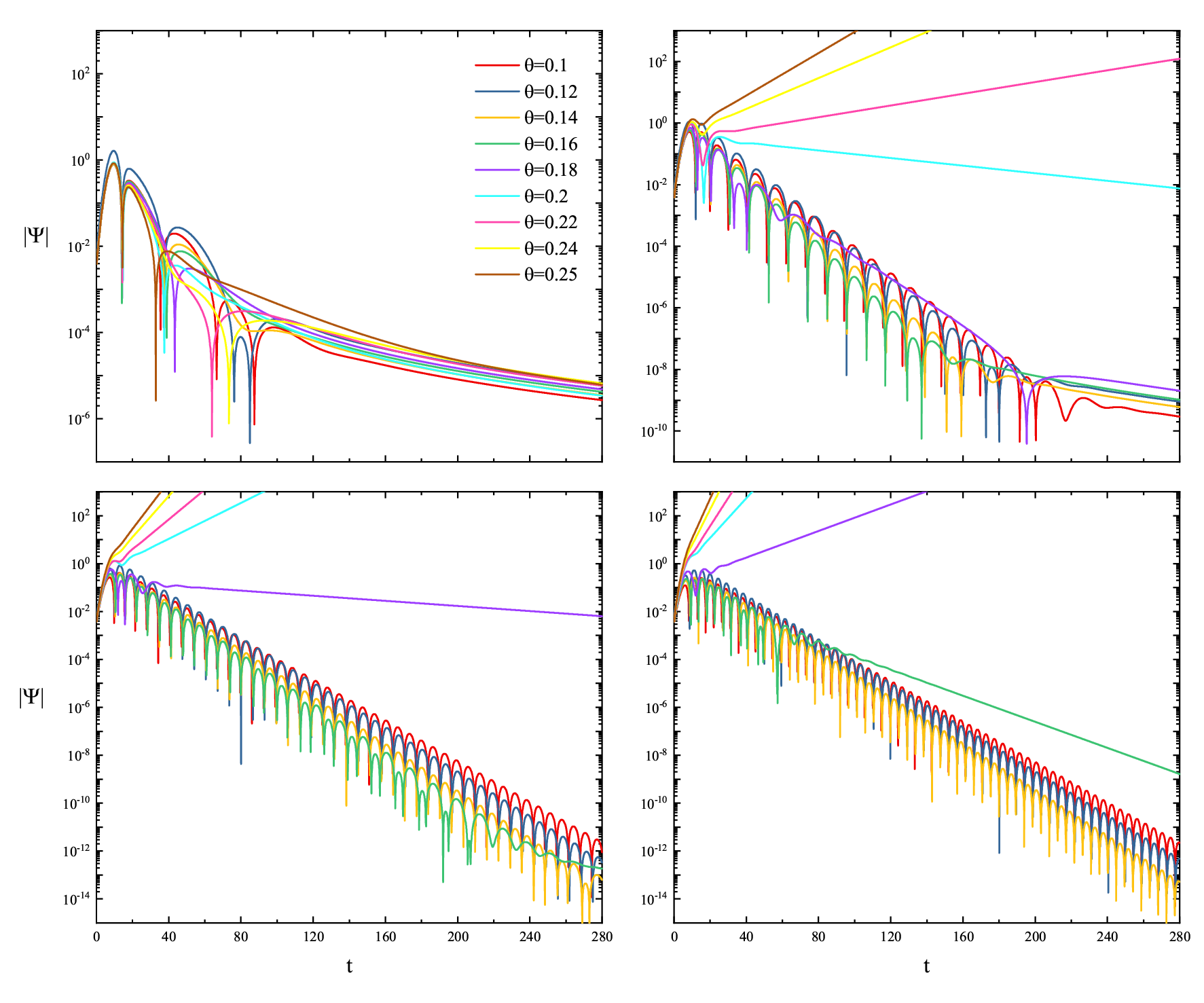}
\caption{Time-domain profiles of a scalar field coupling to Einstein's tensor corresponding to different $\theta$ in the background of non-commutative geometry inspired black hole spacetime. We set $\eta = 20$ in all four panels. The figures from upper-left  to lower-right are corresponding to  $l$ = 0, 1, 2, and 3, respectively. The parameters are selected as $M=1$, $m=0$, $\sigma=3$, $v_{c}=10$.}
\end{figure}

In addition, we compare these numerical results obtained by the time-domain integration method with those obtained by the three methods in the previous section. 
Then we find when $\theta$ is not quite small, the scalar field grows with exponential rate as the $\eta$ is larger than the critical value $\eta_c$, which means that the instability occurs in this case.  
The cause is that the large coupling constant reduces the peak value of the potential and a large negative region appears outside the event horizon.
However, the imaginary parts of the QNMs obtained by the WKB method (the Pad\'e approximants) and the Mashhoon method decreases with the increase of $\eta$, but there is a threshold $\eta_c$ for the numerical results obtained by the AIM method. 
So it is found  that the numerical results by the AIM method and the time-domain integration method are closer to the realistic cases.

\subsection{$\eta_c$ as a function of  parameters $\theta$ and $l$}
We know from the above subsection that as long as the $l$ and $\theta$ values are given, there is a threshold $\eta_c$.
In Table. \ref{etac}, we calculate a series of critical values $\eta_c$ corresponding to different $l$ and $\theta$ values.

\begin{table}[tbh]\centering
\caption{The threshold of $\eta_{c}$ corresponding to different $\theta$ and $l$ values. The parameters are selected as $M=1$, $m=0$, $\sigma=3$, $v_{c}=10$.
\vspace{0.3cm}} \label{etac}
\setlength{\tabcolsep}{4mm}{
\begin{tabular}{|c|c|c|c|c|c|c|c|c|c|c|c|}
\hline
$l$ & \multicolumn{9}{|c|}{$\theta$}   \\ \cline{2-10}
 & $0.10$ & $0.12$ & $0.14$ & $0.16$ & $0.18$ & $0.20$ & $0.22$ & $0.24$ & $0.25$ \\
\hline
0 & $\infty$ & $\infty$ & $\infty$ & $\infty$ & $\infty$ & $\infty$ & $\infty$ & $\infty$ & $\infty$ \\
\hline
1 & 966.0 & 291.4 & 118.4 & 60.9 & 38.9 & 22.6 & 16.5 & 11.6 & 9.18 \\
\hline
2 & 515.7 & 154.2 & 63.8  & 33.3 & 21.2 & 13.0 & 9.6  & 6.9  & 5.63 \\
\hline
3 & 377.4 & 112.2 & 47.2  & 25.0 & 16.0 & 10.1 & 7.5  & 5.5  & 4.56 \\
\hline
4 & 311.8 & 93.0  & 39.4  & 20.0 & 13.5 & 8.7  & 6.4  & 4.8  & 4.06 \\
\hline
\end{tabular}}
\end{table}

The  $\eta_c$  as a function of  $l$ and $\theta$ is shown in Fig. 5 and Fig. 6, respectively.
It can be seen from Fig. 5 that for every fixed $\theta$ value, the changing trend of $\eta_c$ with $l$ is very similar and strictly monotonous.
It can also be seen Fig. 6 that the changing trend of $\eta_c$ with $\theta$ for different $l$ values is very similar and strictly monotonous too.
\begin{figure}[htbp]
\centering
\includegraphics[height=15cm,width=13.5cm]{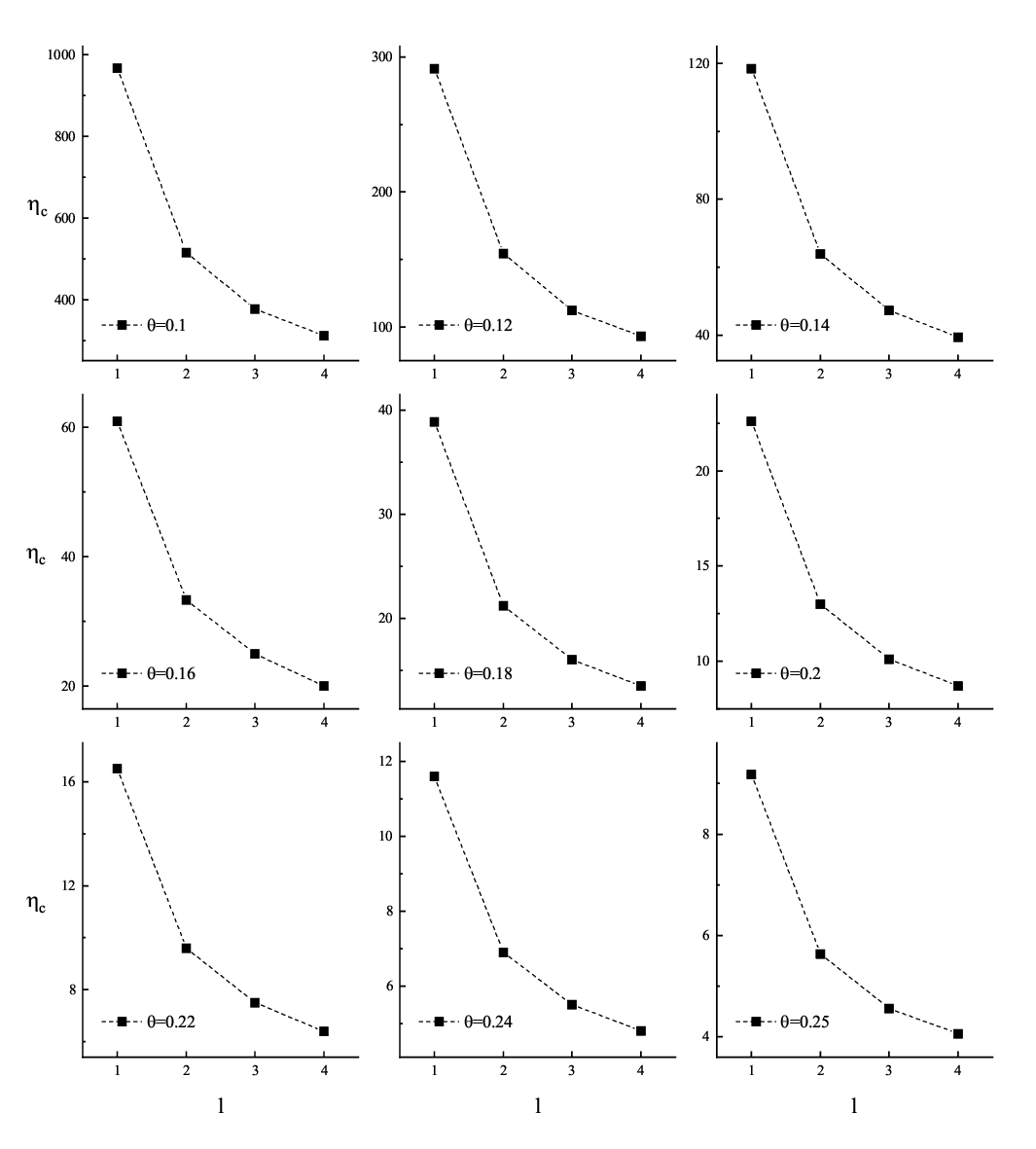}
\caption{The change of $\eta_c$ with $l$ value in the case of different $\theta$.}
\end{figure}

Next, we try to find the fitting function of the $\eta_c$. The fitting method is also used in Ref. \cite{Chen:2010qf} to find the functional relationship between the value of $\eta_c$ and $l$.
Considering that $l$ is a discrete parameter, we think that it is more valuable to use $\theta$ as an independent variable to fit the $\eta_c$ function when the $l$ value is determined.
The numerical algorithm used in the process of function fitting is the Levenberg-Marquardt method in  the \textsl{Origin} software.
We choose $\eta_{c}=a\theta^{b}+c$ as the  function template and get a good fitting effect, as shown in Fig. 6.
In this way, we get the fitting function of $\eta_c$ as a function of $\theta$ corresponding to different $l$ values.

From Fig. 7, we can directly find the changing trend of the fitting parameters $a$, $b$, $c$ with $l$, which leads us to further get the functional relationship between the parameters $a$, $b$, $c$ and variable $l$, respectively. We still use the fitting method to deal with this problem.
In order to make the $R^{2}\rightarrow1$ and the number of parameters obtained by fitting is a little less, we find that a good fitting effect can be obtained by using the power function.

Therefore, we can get the fitting function of $\eta_c(l, \theta)$ as a function of  $l$ and $\theta$:
\begin{equation}
\eta_c(l, \theta) \simeq (n l^{m}) \theta^{(p l^{k})}+(s l^{t}).
\end{equation}
Where $n$, $m$, $p$, $k$, $s$, $t$ are the parameters of the allometric function, respectively. The values and errors obtained by fitting are $n=2.18993\times 10^{-4} \pm 4.72278\times 10^{-5}$, $m=-1.03381 \pm 0.0432$, $p=-6.64059 \pm 0.01147$, $k=0.01041 \pm 0.00181$, $s=12.814 \pm 0.27306$, $t=-0.63754 \pm 0.03176$ respectively.

\begin{figure}[htbp]
\centering
\includegraphics[height=10.9cm,width=16.4cm]{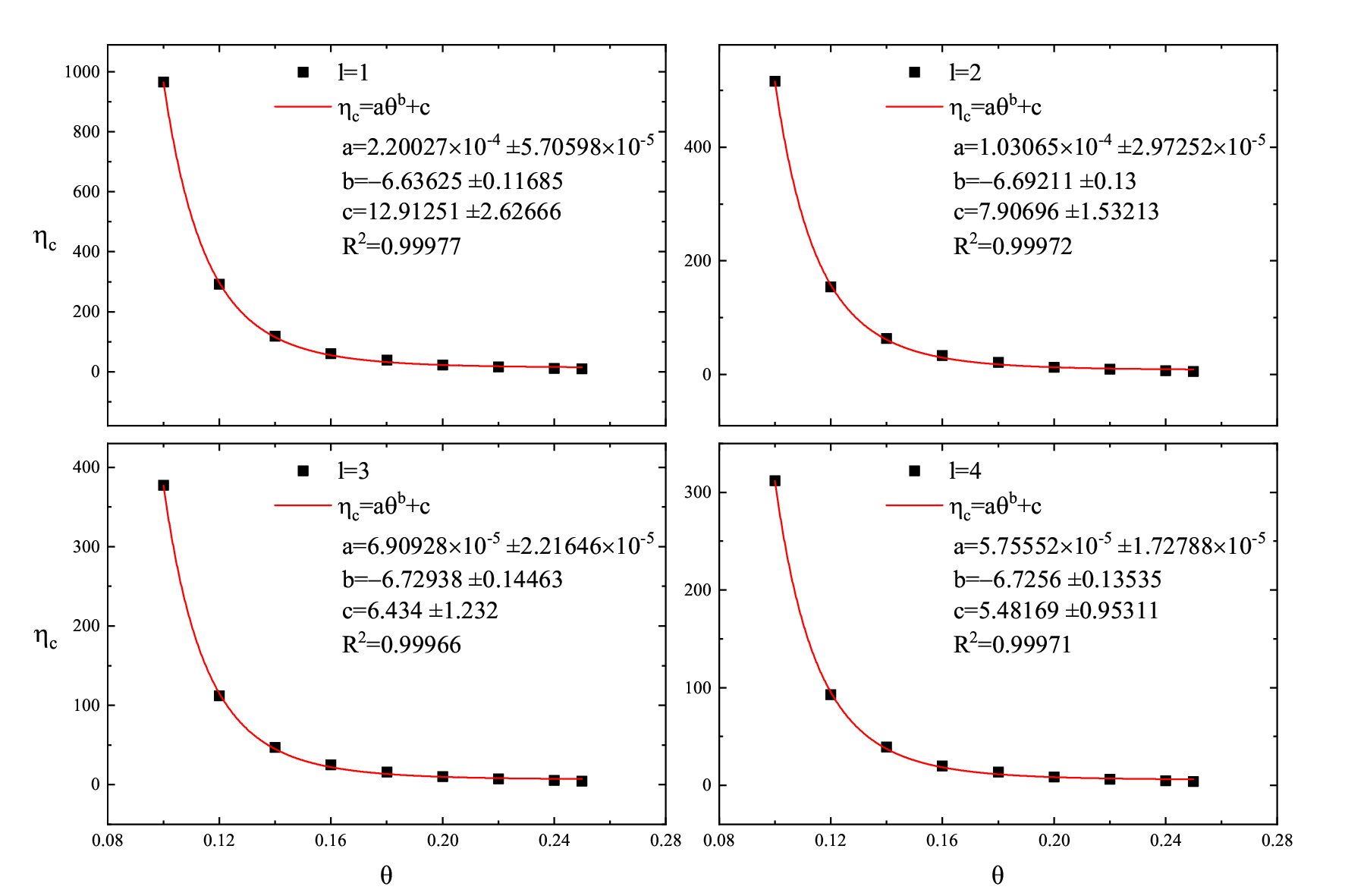}
\caption{The change of $\eta_c$ with $\theta$ value for  different $l$ values. The red solid-line is a fitting function of $\eta_{c}$ with $\theta$ as independent variable. $a$, $b$ and $c$ are fitting parameters and corresponding error values are given.}
\end{figure}

\begin{figure}[htbp]
\centering
\includegraphics[height=5.4cm,width=16.4cm]{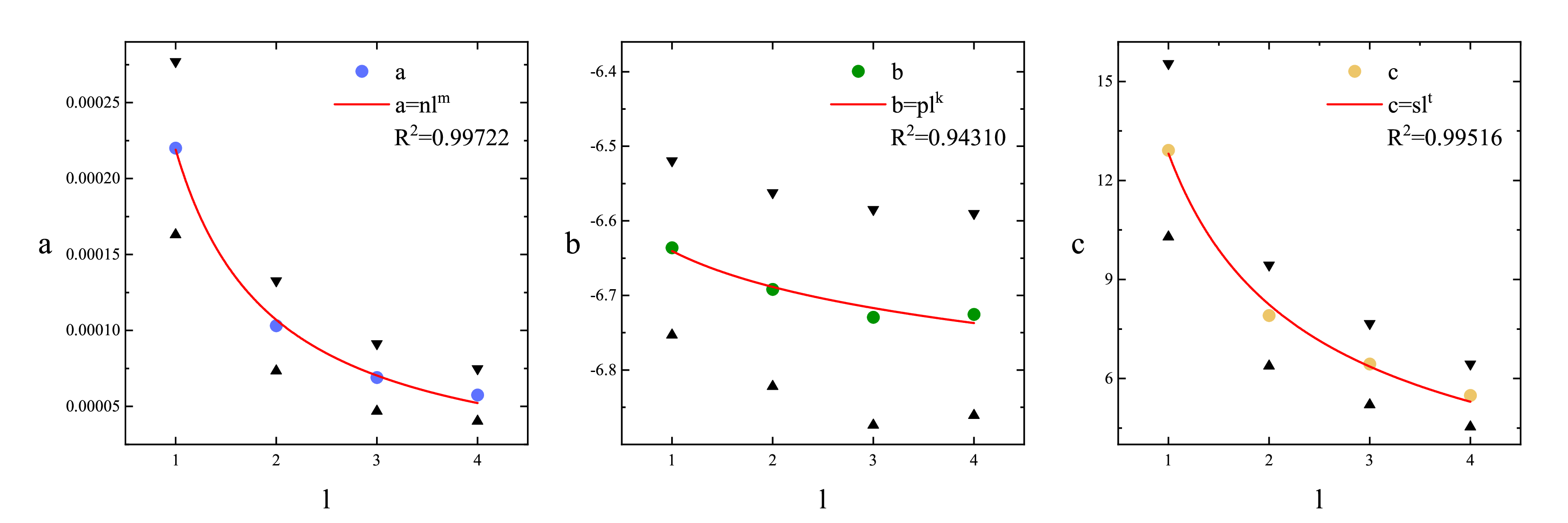}
\caption{From the left panel to the right panel, the red solid-line is the fitting function of the parameters $a$, $b$ and $c$ as a function of $l$, respectively. $\blacktriangledown$ and $\blacktriangle$ correspond to error upper-limit and error lower-limit of $a$, $b$ and $c$, respectively.}
\end{figure}

\section{Summary}
Because the non-commutative black hole is not the solution of the vacuum Einstein's equation, that is, the spherical symmetry $G^{\mu \nu}$ does not disappear.
Therefore, it is  valuable to choose non-commutative black holes to explore the effects of derivative coupling.
In this work, we have studied the QNMs of a scalar field coupled to Einstein's tensor in the background of non-commutative geometry inspired black hole spacetime.
We give the derivation procedure of the motion equation of the scalar field coupled to Einstein's tensor in spherically symmetric spacetime and show that the potential equation of positive coupling parameter will appear singularity in this background, so this paper only discusses the case of negative coupling parameter.

By comparing the numerical results of the WKB method (the Pad\'e approximants), the Mashhoon method and the AIM method, it is found that when the $\theta$ is small, the results of  three methods are in good concordance with the increase of the $\eta$;  when the $\theta$ is larger and the $\eta$ is small, the results of the three methods are almost same too.
However, when the $\theta$ and the $\eta$ is large at the same time, the numerical results by the WKB method (the Pad\'e approximants) and the Mashhoon method are obviously different from those of the AIM method.
Considering the characteristic of the WKB method (the Pad\'e approximants) and the Mashhoon method, the numerical results given by these two methods are not close to the exact values when parameters $\theta$ and $\eta$ are large.

When $\eta=0$,  after comparing the $V_{PT}$ with the effective potential $V_{eff}$, Ref. \cite{Batic:2019zya} considered that the Mashhoon method is suitable for calculating the QNMs in the non-commutative  black hole spacetime.
The increase of the coupling constant $\eta$ affects the shape of the potential of Fig. 1 in tortoise coordinates. This may be the reason for the gradual decline of the calculation accuracy of the Mashhoon method.
On the other hand, the WKB method still fails in these same problem when the $\eta=0$.
The calculations in Ref. \cite{Yu:2018zqd} are consistent with our results (See Fig. 2), which means that the WKB  approximation is not suitable for the non-commutative black hole spacetime. It is reported that for different black hole spacetimes, the high-order WKB calculation results  can not guarantee convergence.

This observation also shows that when parameters $\theta$ and $\eta$ are larger together, the numerical results obtained by the WKB method (the Pad\'e approximants) and the Mashhoon method are not accurate.  
The time-domain integration method is used to show there exist a threshold $\eta_{c}$ defined in the Section 5.A, which is consistent with the numerical results calculated by the AIM method, that is, there will be dynamical instability when $\eta>\eta_c$.
The dynamical instability may be due to the negative region of the effective potential outside the maximum event horizon.
Furthermore, by means of  a numerical fitting,  we obtained that the functional relationship between the threshold $\eta_c$ and the non-commutative parameter $\theta$ satisfies $\eta_{c}=a\theta^{b}+c$ for a fixed $l$ approximately.

At last, it is worth noting that the exact solution of the non-commutative black hole with non-minimally derivative coupling (NC-NMDC) has not been found so far.
However, the solution of the Schwarzschild black hole with non-minimally derivative coupling in asymptotically flat spacetime is given in Refs. \cite{Rinaldi:2012vy, Minamitsuji:2013ura}, which indicate that
 when the $\eta$ is adequate large, this new metric can be reduced to that of the Schwarzschild black hole solution.
So it is reasonable generalization and hypothesis that the metric of the non-commutative black hole with non-minimally derivative coupling  is approximately equal to that of the non-commutative Schwarzschild black hole when the $\eta$ is adequate large.
Furthermore, when the $\eta$ is very large, the conclusion related to the instability of the dynamical evolution of  a scalar field in this background is reliable.

\begin{acknowledgments}
The author thanks Dr. X. J. Zhang for his positive help and useful discussion. This work was supported by the National Key Research and Develop Program of China under Contract No. 2018YFA0404404 and  the Key  Research Program of the Chinese Academy of Sciences (Grant No. XDPB09-02).
\end{acknowledgments}

\bibliography{NC-NMDC-Ref0309}

\end{document}